\documentclass[useAMS, usenatbib]{mnras}
\usepackage[T1]{fontenc}
\usepackage{ae,aecompl} 
\usepackage{amsmath}    
\usepackage{amssymb}    
\usepackage[pdftex]{graphicx}
\title[Spectroscopy of AG Dra]{Spectroscopic view on the outburst activity of the symbiotic binary AG
Draconis}
\author[L. Leedj{\"a}rv et al.]
{L. Leedj{\"a}rv,$^1$\thanks{E-mail: laurits.leedjarv@to.ee}
  R. G{\'a}lis,$^2$ L. Hric,$^3$ J. Merc$^2$ and M. Burmeister$^1$ \\
$^1$Tartu Observatory, Observatooriumi 1, T{\~o}ravere, 61602
Tartumaa, Estonia \\
$^2$Faculty of Science, P. J. {\v S}af{\'a}rik University, Park Angelinum 9,
Ko{\v s}ice 040 01, The Slovak Republic \\
$^3$Astronomical Institute, Slovak Academy of Sciences, Tatransk{\'a}
Lomnica 059 60, The Slovak Republic} 

\date{Accepted 2015 November 27. Received 2015 November 27; in original form
2015 September 14}

\pubyear{2015}

\begin{document}
\label{firstpage}
\pagerange{\pageref{firstpage}--\pageref{lastpage}}
\maketitle

\begin{abstract}
Variations of the emission lines in the spectrum of the yellow symbiotic
star AG Dra have been studied for over 14 years (1997--2011), using more
than 500 spectra obtained on the 1.5-metre telescope at Tartu Observatory, Estonia. The
time interval covered includes the major (cool) outburst of AG Dra that
started in 2006.  Main findings can be summarized as follows: (i) cool and
hot outbursts of AG Dra can be distinguished from the variations of optical
emission lines; (ii) the Raman scattered emission line of O\,{\sc vi} at $\lambda$ 6825
almost disappeared during the cool outburst; (iii) lower excitation emission
lines did not change significantly during the cool outburst, but they vary in
hot outbursts and also follow orbital motion; (iv) similarity of variations
in AG Dra to those in the prototypical symbiotic star Z And allows to suggest that a
"combination nova" model proposed for the latter object might also be responsible
for the outburst behaviour of AG Dra. 

\end{abstract}


\section{Introduction}
\label{sect1}

\mbox{AG Dra} belongs to the subclass of so-called yellow symbiotic stars. Its cool
component is of early K spectral class instead of M as in most other
symbiotic stars. The hot component is a white dwarf 
with an effective temperature of
100\, 000 K and a luminosity of $\sim 10^3 L_{\sun}$ \citep{mik95} obtained
from the IUE data. However, \citet{sion12} have derived $T_{\rmn{eff}} =
80\,000$ K from the FUSE far-UV spectrum. The orbital period of the
binary system is $550^{\rmn d}$ \citep{mein79, smith96, fekel00}.
A shorter period around $350-380$ days has also been
detected in photometric and spectroscopic variability of \mbox{AG Dra}
\citep{bast98, gal99, fried03}. \citet{gal99} ascribed this period to the
pulsations of the K giant. 

By its photometric variability, \mbox{AG Dra} can be classified as a classical
symbiotic star of \mbox{Z And} type. Its light curve, available since 1890
\citep{rob69}, shows frequent brightenings of about $1-1.4$
magnitudes in the $V$/visual band and up to 2.3 and 3.6 magnitudes in the $B$ and $U$
bands, respectively. Major outbursts occur in intervals of $12-15$ years (in 1936,
1951, 1966, 1980, 1994, 2006), and are usually followed by minor scale
outbursts in intervals of about one year. \citet{gonr99} have distinguished
between cool and hot
outbursts of \mbox{AG Dra}. Major outbursts in the beginning of active phases are
usually cool ones, during which the expanding pseudo-atmosphere of the white dwarf cools down and
the \mbox{He\,{\sc ii}} Zanstra temperature drops. In smaller scale hot outbursts,
the \mbox{He\,{\sc ii}} Zanstra temperature increases or remains unchanged.

The outburst activity of \mbox{AG Dra} over 120 years was studied in our recent paper
\citep{hric14}, using mostly photometric observations. There are numerous
other studies available, making use of optical spectroscopy as well as other
methods of observations 
\citep[see e.g.][and references therein]{tomtom02, leed04, gonr08, mun09,
sho10}.
However, a lot of open questions remain concerning 
the nature and physical mechanism of the outbursts in particular. In the present
paper, we study the variability of emission lines in the optical
spectrum of \mbox{AG Dra} during almost 14 years (September 1997 to April 2011).
This time interval includes the major outburst that started in July 2006. A
brief description of the behaviour of \mbox{AG Dra} around this event was given
by \citet{leedbur12}. In the
present paper, we undertake an analysis of \mbox{AG Dra} over a longer timeframe, and
attempt to compare it with other symbiotic stars. Section \ref{sect2} describes the
observations. In Sect. \ref{sect3} we present the main features of variability of the
strongest optical emission lines. In Sect. \ref{sect4} we attempt to find 
similarities of \mbox{AG Dra} to the prototypical symbiotic
star \mbox{Z And} and other symbiotic stars. Discussion and conclusions are
given in Sect. \ref{sect5}.

\section{Observations and data reduction}
\label{sect2}

Spectroscopic observations of \mbox{AG Dra} used in the present paper have been
carried out on the 1.5-metre telescope at Tartu Observatory, Estonia.
The equipment and methods of observations and their reductions are described
by \citet{leed04} (hereafter L04). 
Starting from March 2006, a new Andor
Newton DU-970N CCD camera with $400 \times 1600$ pixels (pixel size $16
\times 16 {\mu}\rmn{m}$) has been used. With this camera, dispersions $\sim0.47$ {\AA} pix$^{-1}$ 
or $\sim0.16$ {\AA} pix$^{-1}$ in the red spectral
region, and $\sim0.57$ {\AA} pix$^{-1}$ in the blue region were achieved. Test measurements have shown that
the change of CCD detectors has no systematic effect on the obtained results.

Altogether 515 spectra of \mbox{AG Dra}, recorded between September 1997 and April
2011, are used in the present paper: 196 in the blue (H\,$\beta$) spectral
region, 221 in the red (H\,$\alpha$) spectral region with lower dispersion,
and 98 in the red region with higher dispersion. A few spectra have
been sporadically recorded after April 2011, when the star was continuously in the
quiescent stage Q6 (see Sect. \ref{sect3}), data from those spectra do not
affect our analysis. Reductions of the spectra
were done with the help of the ESO {\sc midas} software package, applying the
standard procedures of subtracting the bias and sky background, calibrating
into the wavelength scale, and normalizing to the continuum.

We focus on the strongest emission lines in the wavelength regions
under study: the hydrogen Balmer lines H\,$\alpha$ ($\lambda$ 6563) and
H\,$\beta$ ($\lambda$ 4861), the neutral helium \mbox{He\,{\sc i}} line at $\lambda$
6678, the ionized helium \mbox{He\,{\sc ii}} line at $\lambda$ 4686, and the Raman
scattered \mbox{O\,{\sc vi}} line at $\lambda$ 6825. Equivalent widths (EW), peak
intensities relative to the continuum, and positions of these lines were
measured. 
Uncertainties of the EWs were estimated from repeated test measurements of a
few spectra, they range from about 3 per cent for the strongest lines
to about 10 per cent for weak lines.
 
\begin{figure*}
\includegraphics[width=168mm]{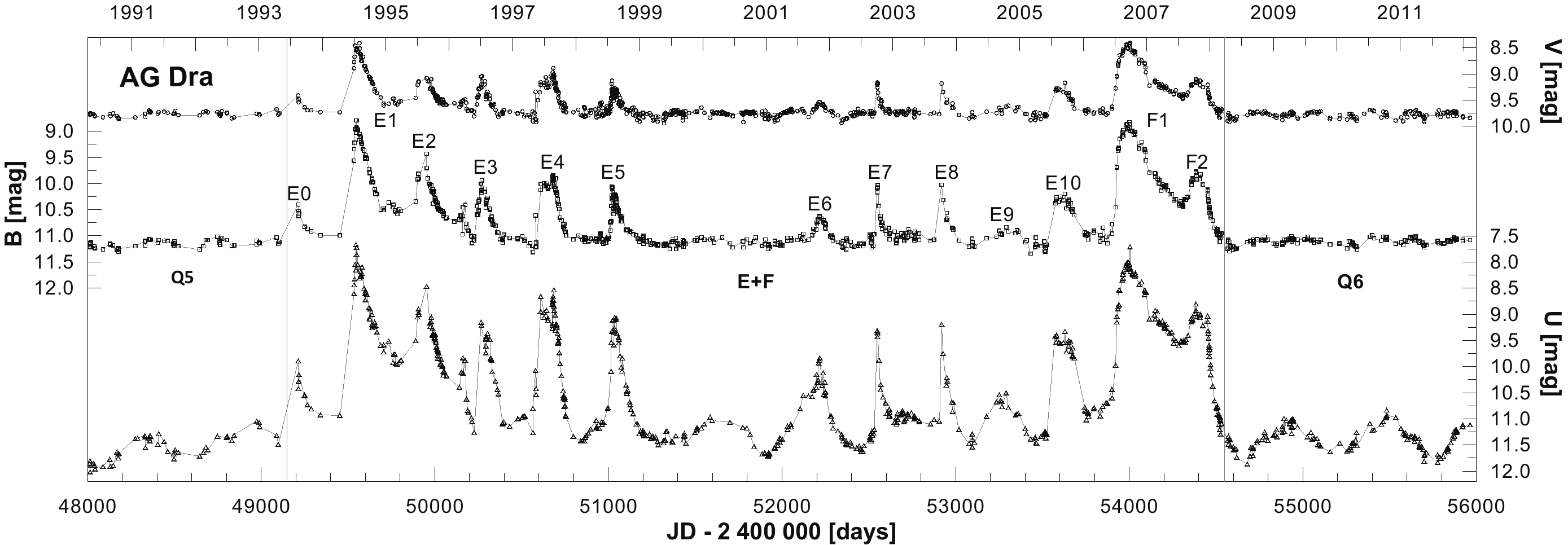}
\caption{$U$ (bottom), $B$ (medium), and $V$ (top) light curves of \mbox{AG
Dra} from 1990 to 2012.
Active stages E and F are confined between the vertical lines, with individual
outbursts identified. The thin curves show a spline fit to the data points.}
\label{fig1}
\end{figure*}

EWs may not be a good indicator
of emission line variations when the underlying continuum is variable.
Therefore, we have calculated absolute fluxes emitted in the lines, using the
same photometric data as in \citet{hric14}. In addition, data from
\citet{skop07} were applied. For interstellar reddening the value $E(B-V)=0.05$ was
adopted according to \citet{mik95}. Interpolation of the photometric data in
wavelength scale and in time brings along additional errors, and so
uncertainties in the emission line absolute fluxes may reach up to 15--20
for the weakest lines.

\section{Characteristics of the emission lines from 1997 to 2011}
\label{sect3}
\subsection{Variability and correlations with the brightness}

In Fig. \ref{fig1} we present the $UBV$ light
curves of \mbox{AG Dra} from 1990 to 2012, consisting of the quiescent and
active stages. Individual events are marked, and these labels will be
referred to throughout the paper.
The EWs of the emission lines in the spectrum of \mbox{AG Dra}
from April 2003 to April 2011 are presented in Table \ref{table1} (full table is only
available online). In Table
\ref{table1}, we have
applied the frequently used convention in which EWs of emission lines
are given as positive numbers.

\begin{table}
\caption{Equivalent widths of the strongest emission lines in
the spectrum of \mbox{AG Dra} from 2003 to 2011. Only the first few lines are
shown. The full table is available online. Earlier data
from September 1997 to March 2003 are available at
http://cdsweb.u-strasbg.fr/cgi-bin/qcat?J/A+A/415/273).} 
\label{table1} 
\centering
\begin{tabular}{r c c c c c}
\hline
JD & \multicolumn{5}{c}{EW({\AA})} \\
$-2\,400\,000$ & H\,$\alpha$ & H\,$\beta$ & \mbox{He\,{\sc i}} 6678 &
\mbox{He\,{\sc ii}} 4686 & \mbox{O\,{\sc vi}} 6825 \\
\hline
52\,731.410 & 112.8 & 33.1 & 2.20 & 15.85 & 7.07 \\
52\,748.434 & 104.4 & 36.0 & 2.01 & 17.54 & 6.39 \\
52\,749.422 & - & 29.6 & 1.98 & 17.24 & 6.90 \\
52\,751.414 & 105.6 & 31.2 & 1.88 & 17.29 & 6.57 \\
52\,755.387 & 121.0 & - & 1.94 & - & - \\
52\,756.338 & 112.7 & - & 2.08 & - & - \\
52\,762.387 & 103.1 & 32.0 & 1.68 & 17.00 & 6.02 \\
\hline
\end{tabular}
\end{table}

\begin{figure*}
\includegraphics[width=168mm]{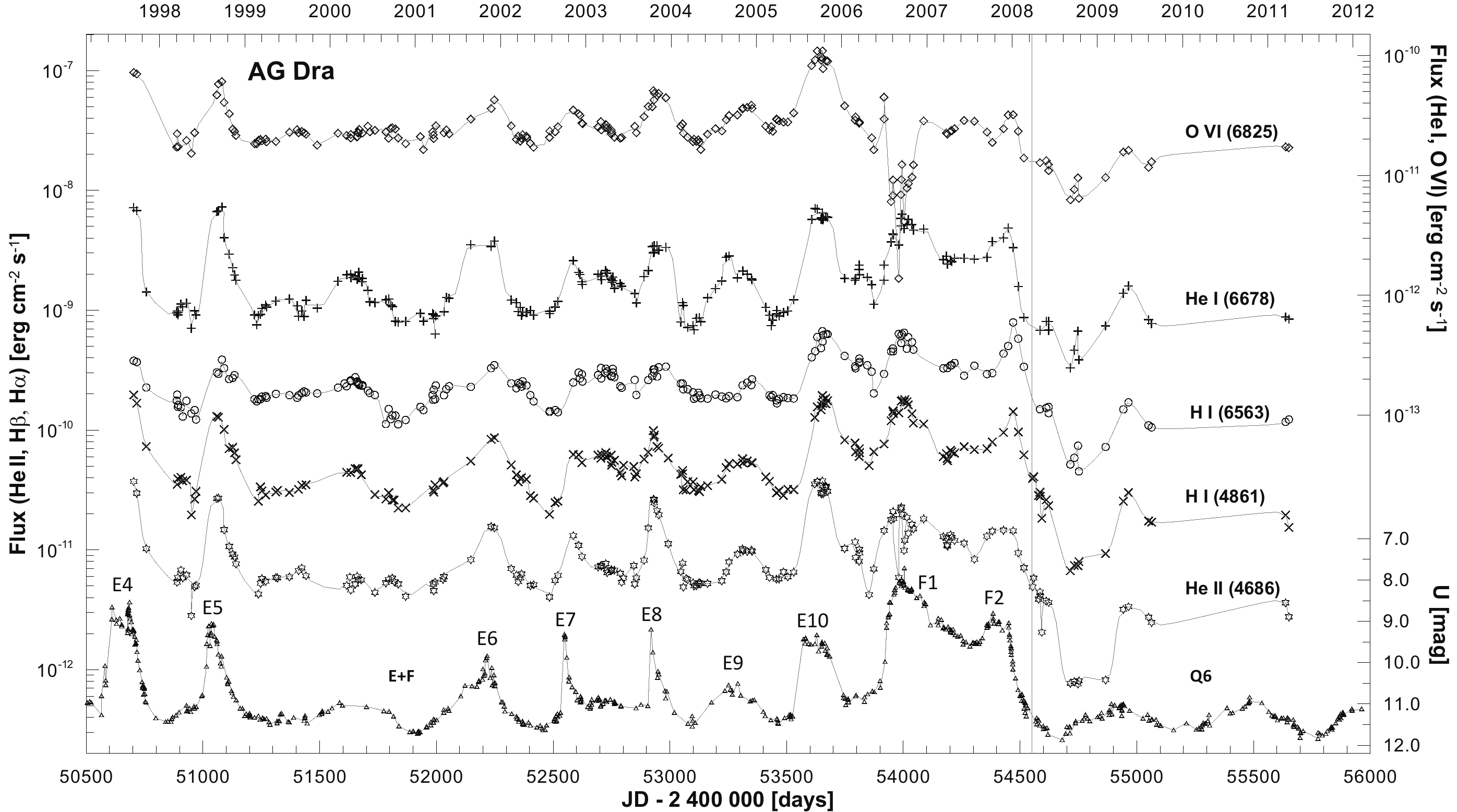}
\caption{Absolute fluxes in the strongest emission lines of \mbox{AG Dra} together with the
$U$ light curve. Thin lines show a spline fit to the data points.}
\label{fig2}
\end{figure*}

In \citet{hric14} we presented variability of the EWs together with the $U$
light curve in fig. 8. 
Figure \ref{fig2} shows the temporal variations of the absolute fluxes of the same emission lines.
One remarkable feature of these variations is the brightening of all the lines 
during the outburst E10 in 2005. 
Even more noticeable is the deep minimum in the fluxes 
around JD\,2\,454\,800 (2008--2009), after a major double-peaked outburst
(F1, F2).
At the same time, there are  no peculiarities in the
light curves; the $U$, $B$ and $V$ magnitudes correspond to their usual values in quiescence.
The major (cool) outburst of \mbox{AG Dra} that started after
JD\,2\,453\,900 (July 2006, F1), 
is not specifically distinct in the fluxes of
hydrogen and helium lines, but a weakening of the Raman scattered
O\,{\sc vi} line is clearly visible. Such a disappearance
of the high excitation emission line in the spectrum of \mbox{AG Dra}
during a major outburst was also observed by \citet{mun09} and
\citet{sho10}. \citet{burleed07} detected a similar weakening
of the O\,{\sc vi} $\lambda$ 6825 line in the spectrum of the
prototypical symbiotic star \mbox{Z And} in summer 2006,
when the star underwent a strong outburst accompanied by the
ejection of bipolar jets. In the case of \mbox{Z And}, the He\,{\sc
ii} $\lambda$ 4686 line also practically disappeared at the
beginning of the outburst, while it did not display remarkable
variations during the 2006 outburst of \mbox{AG Dra}.

\begin{figure}
\includegraphics[width=82mm]{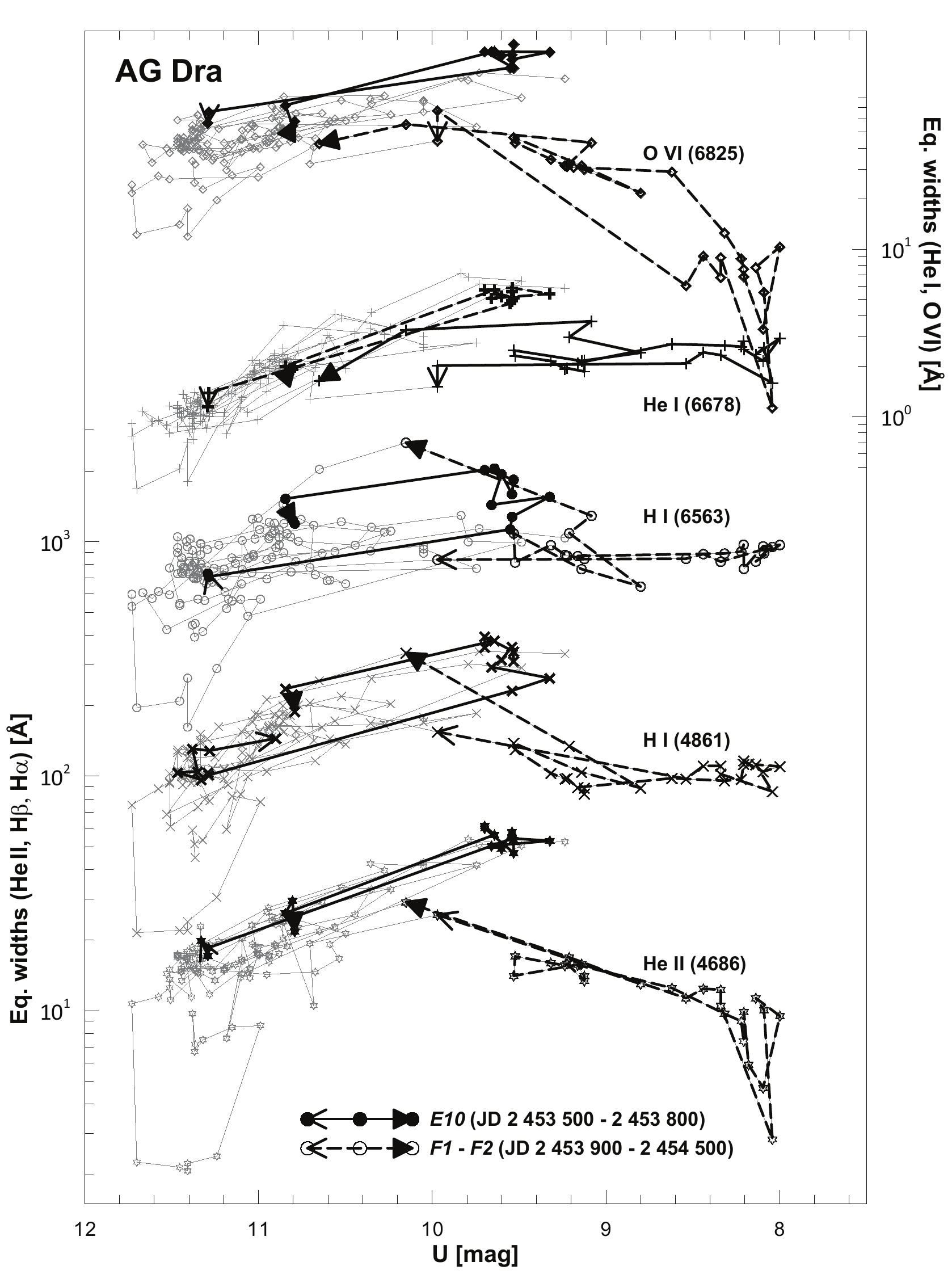}
\vspace{1mm}
\caption{EWs of the strongest emission lines in logarithmic scale vs. $U$ magnitude. Data points 
are connected chronologically. The thick solid line connects the
points that correspond to the outburst episode E10, and the thick dashed line
connects those from the major outbursts F1 and F2.}
\label{fig3}
\end{figure}

In L04, we found a
more or less linear correlation between the $U$ magnitude and
logarithms of the EWs (fig. 2 in L04). 
In Fig. \ref{fig3} we present (in logarithmic scale) the dependence of the 
EWs of the emission lines on the $U$ magnitude over a longer
time span, including the cool outburst in 2006. Chronological connection of
the points enables us to see the different behaviour of the EWs during hot and cool
outbursts. 
We have also studied the dependence of the EWs on the $B$ magnitude. 
Qualitatively they are almost congruent to those in Fig. \ref{fig3}.
The EWs of all lines increase with the brightening in $U (B)$, until $U (B)$ reaches
a value of $9.5-9.4 (10.2-10.0)$. A different behaviour is seen at brighter magnitudes. 
The EWs of H\,$\alpha$ and He\,{\sc i} 6678 remain more or less
constant.
In H\,$\beta$, 
a light decrease is seen for brighter $U$ magnitudes.
Finally, the high excitation He\,{\sc ii} and O\,{\sc vi} lines show
a remarkable decline.
Hence, the magnitudes $U \sim 9.4$ and $B \sim 10.2$ apparently mark the
brightness limits above which the onset of a cool outburst of the hot
component of \mbox{AG Dra} can be expected.

\subsection{Ratio of the He\,{\sc ii} $\lambda$ 4686 and H\,$\beta$ lines}

\begin{figure}
\includegraphics[width=84mm,angle=0]{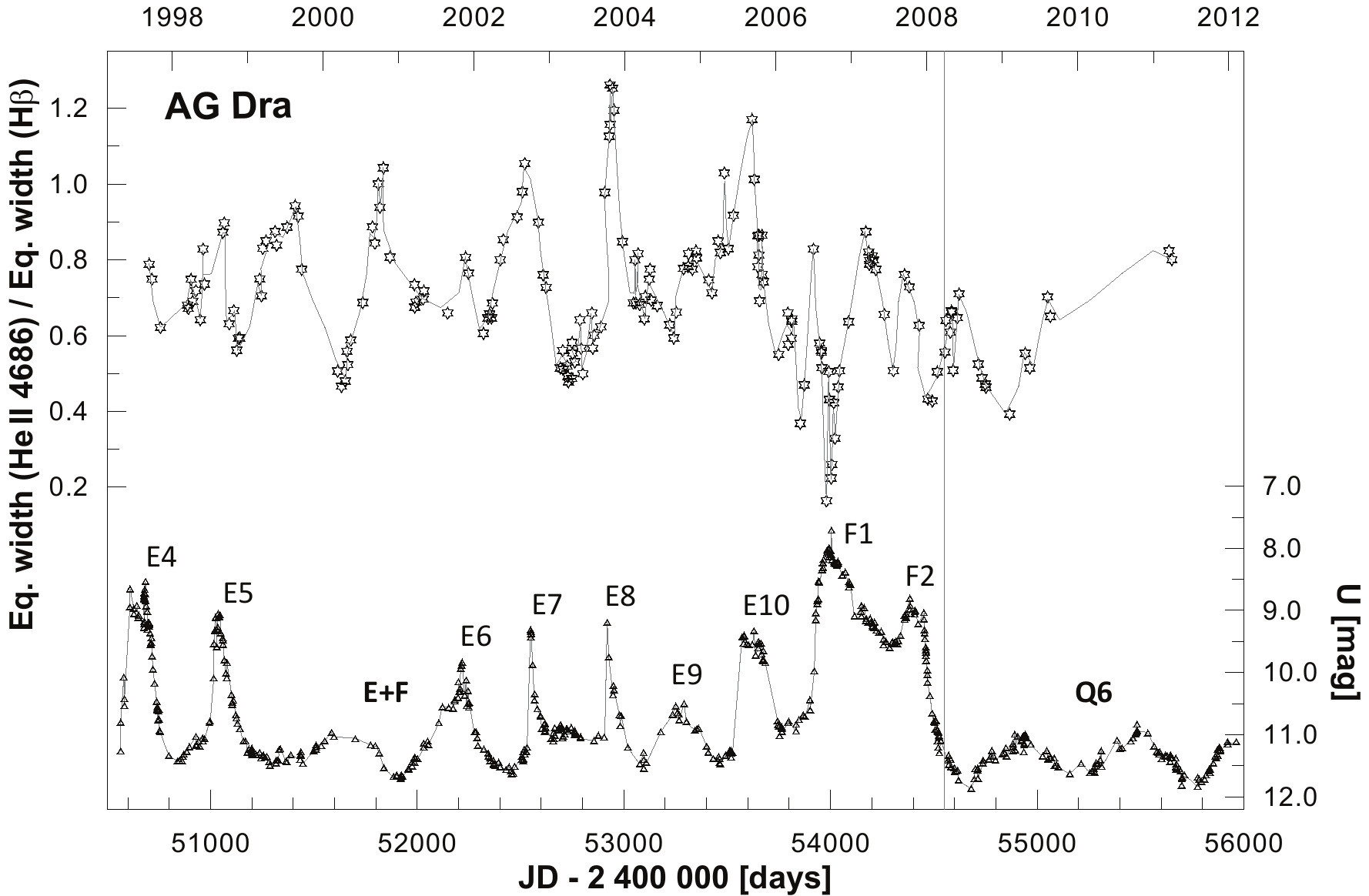}
\caption{Ratio of the EWs of the two strong emission lines, He\,{\sc ii}
$\lambda$ 4686 and H\,$\beta$, in time.}
\label{fig4}
\end{figure}

In L04 we reported that the average value of the EW ratio of the ionized
helium line at $\lambda$ 4686 and the H\,$\beta$ line (He\,{\sc
ii}/H\,$\beta$) was $0.79 \pm 0.15$ between 1997 and 2003, in accordance with
\citet{gonr99}. There has only been one short episode 
when this ratio slightly exceeded 1.0. \citet{sho10} confirm this result,
but they have detected no epoch with He\,{\sc ii}/H\,$\beta > 1.0$.
Our measurements over the time interval from 1997 to 2011 are shown in
Fig.~\ref{fig4}. 
During four episodes the ratio marginally exceeds 1.0.
A significant decrease
to $ \sim 0.2$ can be detected around JD\,2\,454\,000 (major outburst F1), 
and there are a few more occasions when the ratio drops to 0.5 or slightly
less. Due to these low values the long-term average of the line ratio over
the period 1997--2011 was $0.70 \pm 0.18$.

The He\,{\sc ii}/H\,$\beta$ ratio can actually be considered as a proxy to
the temperature of the hot component. \citet{iij81} has provided a simple
formula for assessment of the effective temperature of a central source of
ionizing photons from the nebular emission-line fluxes:

\begin{equation}
T_{\rmn{hot}} (\rmn{in\,10^4\,K}) = 19.38 \sqrt{2.22F_{\rmn{4686}} \over (4.16F_{\rmn{H\,\beta}} +
9.94F_{\rmn{4471}})} + 5.13, 
\end{equation}
where $F_{\rmn{4686}}, F_{\rmn{H\,\beta}},$ and $F_{\rmn{4471}}$ are the
fluxes in the emission lines of He\,{\sc ii} $\lambda$ 4686, H\,$\beta$, and
He\,{\sc i} $\lambda$ 4471, respectively. \citet{soko06} have neglected the
flux of the He\,{\sc i} $\lambda$ 4471 line in the case of \mbox{Z And}, because
$F_{\rmn{4471}} \la 0.1 F_{\rmn{H\,\beta}}$. Approximately the same relation
is valid for \mbox{AG Dra}. With a similar simplification, the EW ratio
can be used instead of the flux ratio. So, we obtain:

\begin{equation}
T_{\rmn{hot}} (\rmn{in\,10^4\,K}) \approx 14.16 \sqrt{
\rmn{EW}_{\rmn{4686}} \over \rmn{EW}_{\rmn{H\,\beta}} } + 5.13.
\end{equation}

Equation (2) yields a temperature of the hot component at about
170\,000 K for the average line ratio of 0.70 as well as  
maximum (210\,000 K) and minimum (108\,000 K) values for the extrema 1.26
and 0.16, respectively. These temperatures are somewhat higher than those
derived by \citet{gonr99}, and in particular that by \citet{sion12}.
We are aware that by neglecting the
He\,{\sc i} line the temperature increases. At the same time,  
assumption of a radiation-bounded nebula made in the derivation of Eq. (1) may not be
fully valid in
the case of \mbox{AG Dra}. Nevertheless, Eq. (2) provides a useful
indication of the temperature changes, and Fig. \ref{fig4} can hence be
regarded as representative for the temporal evolution of the hot component's temperature
variations. The temperature decrease during the major (cool)
outburst F1 around JD 2\,454\,000 is clearly seen. Other episodes of
low He\,{\sc ii}/H\,$\beta$ values center around JD 2\,454\,480 (F2)
and 2\,454\,840 (quiescence Q6). The latter episode coincides with the detection of the
lowest flux values in both low- and high-excitation emission
lines as presented in Fig. \ref{fig2}. Similar drops 
to $\sim 0.5$ can also be seen around JD 2\,451\,640 (possible quiescent
episode between E5 and E6) and 2\,452\,750 (between E7 and E8). Some weakening of the
Balmer and He\,{\sc i} emission lines at the same times can be
detected, but these are not comparable to the event at JD
2\,454\,840.

One can also notice a few exceptionally high He\,{\sc ii}/H\,$\beta$ values,
$ > 1.0$ around JD 2\,451\,810 and 2\,452\,520, and about 1.2 around JD
2\,452\,940 and 2\,453\,620, respectively. Interestingly, three 
of those episodes can be associated with hot outburst type brightenings (E7,
E8 and E10) were the $U$ and $B$ magnitudes reach the turning point in
Fig. \ref{fig3}, while the first event took place during a short quiescence phase between the outbursts
E5 and E6. 

\subsection{The O\,{\sc vi} $\lambda$ 6825 line}

\begin{figure}
\includegraphics[width=84mm,angle=0]{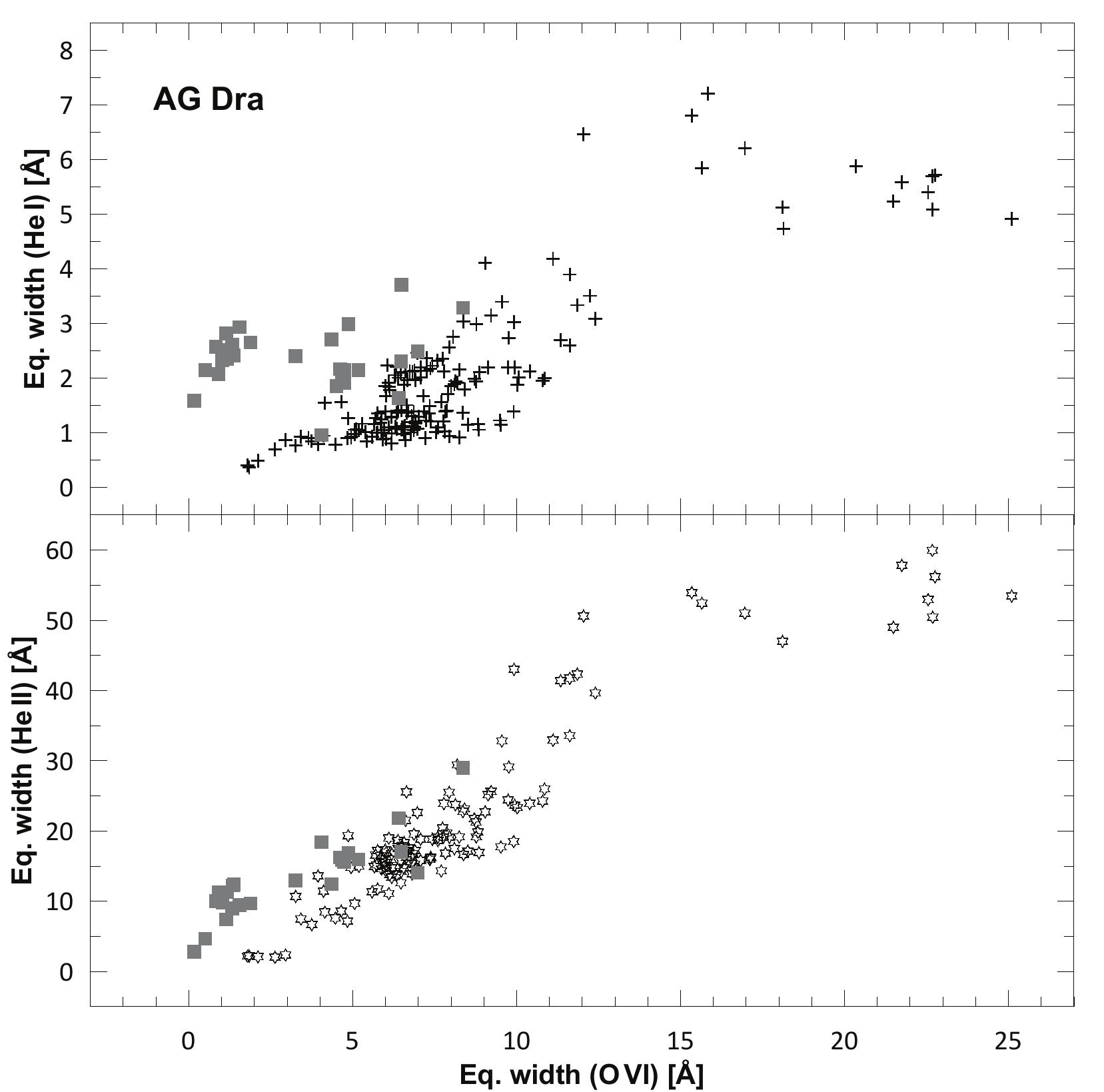}
\caption{Dependence of the EWs of the neutral (He\,{\sc i} $\lambda$ 6678,
top) and ionized (He\,{\sc ii} $\lambda$ 4686, bottom) helium emission lines on the EWs of
the Raman scattered O\,{\sc vi} $\lambda$ 6825 line. Grey squares denote
the points from the cool outburst (F1+F2).}
\label{fig5}
\end{figure}

Broad emission lines in the spectra of symbiotic stars at about 6830 and 7080
\AA~
remained a mystery, until \citet{sch89} identified them as a product of Raman
scattering of photons from the \ion{O}{vi} resonance lines at 1032 and
1038 \AA~ at atoms of neutral hydrogen. The formation of these lines
requires specific physical conditions -- simultaneous presence of a 
hot radiation source, capable of ionizing oxygen atoms five times, and
enough
neutral hydrogen atoms -- which are met almost exclusively in symbiotic stars.
\citet{arrtp03} have identified the same lines in the spectra of a few
planetary nebulae where also broad wings of the H\,$\alpha$ line are
supposed to form in a similar process involving the Ly\,$\beta$ photons
\citep{leehy00}.  Other Raman scattering processes can be expected, for
example, the line at 4881 \AA~originating from the Ne\,{\sc vii} $\lambda$
973 photons, was recently identified in the spectrum of the Galactic symbiotic star
\mbox{V1016 Cyg} \citep{lee14} and of symbiotic star candidates in the galaxy \mbox{NGC
205} \citep{gonc15}. 
Very intruiging is the discovery of the
Raman scattered O\,{\sc vi} lines by \citet{torr12} in the spectrum of the
massive luminous B[e] star \mbox{LHA 115-S 18} in the SMC. 
\citet{torr12} have
proposed two alternative explanations for the appearance of those spectral
features: \mbox{LHA 115-S 18} could be (1) a B[e] supergiant in a pair with a hot main-sequence star accreting
mass, or (2) a Luminous Blue Variable (LBV) object. 

\begin{figure}
\includegraphics[width=84mm,angle=0]{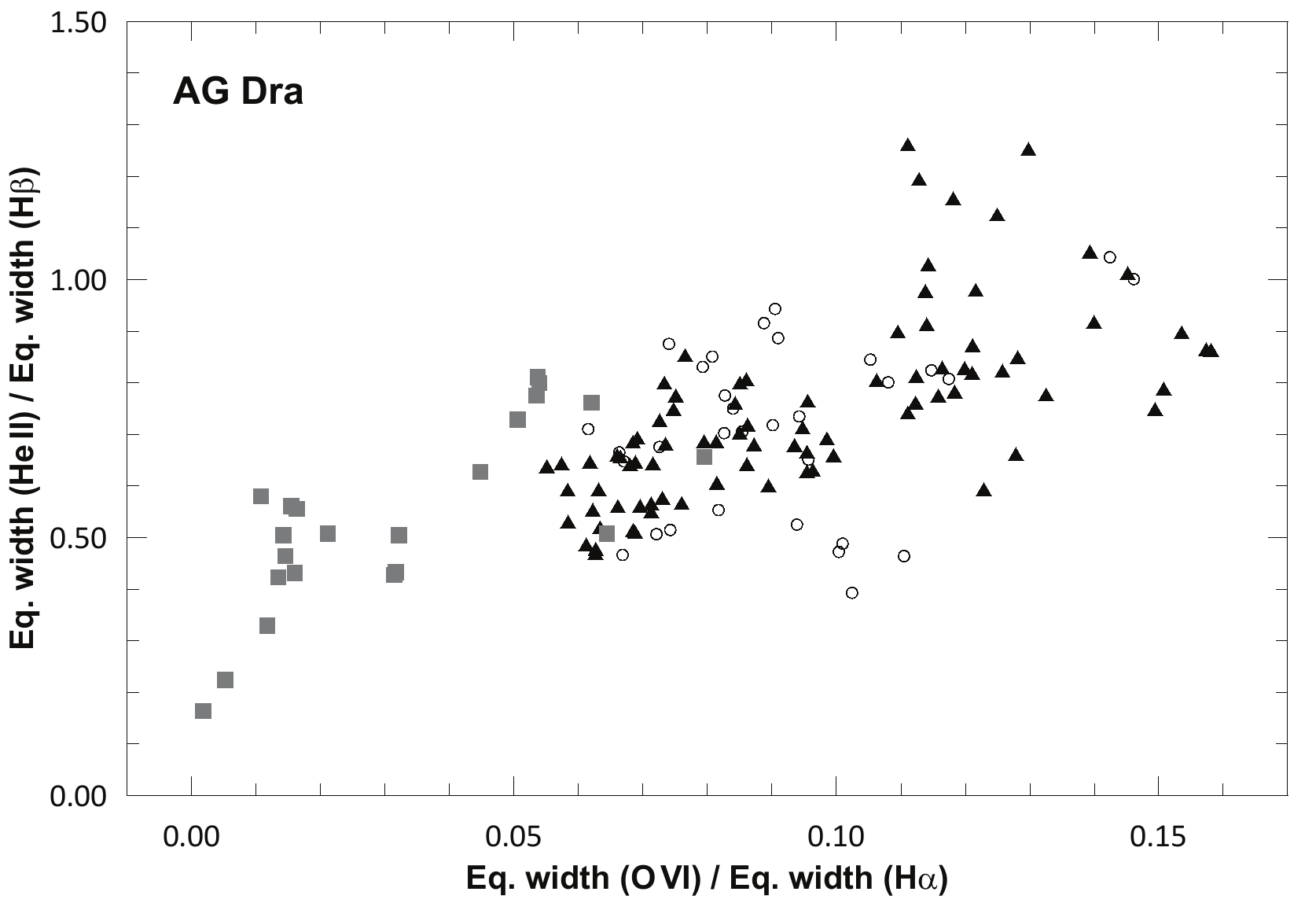}
\null
\caption{Relations of the EWs. Open circles mark data points in quiescence
and triangles in hot outbursts. (Grey) squares denote the points from the cool
outburst (F1+F2).}
\label{fig6}
\end{figure}

Raman scattered O\,{\sc vi} lines are always present in the
spectrum of \mbox{AG Dra}, although they almost disappeared during the
cool outburst in 2006 \citep{mun09,sho10}. Our observations
confirm this finding. 
Fig. \ref{fig5} shows the relation between the EWs of the helium and the O\,{\sc vi} $\lambda$ 6825
lines. With more data points and wider ranges of the EW
values than those in \citet{sho10}, the correlation between the
He\,{\sc ii} and O\,{\sc vi} lines remains more or less linear.
Some discontinuity and deviations from linearity can
be detected at EW(O\,{\sc vi}) values between 10 and 15
\AA. The He\,{\sc i} line also roughly follows a linear
correlation, but with a bigger scatter, especially at the highest
EW values, and the data points from the cool outburst F1+F2  mostly deviate from a linear
relation. 
A simple interpretation of these relations could be that during the cool
outburst the temperature of
the hot component considerably decreased,
so that the high excitation O\,{\sc vi} and He\,{\sc
ii} lines significantly faded and almost disappeared, but leaving the
lower excitation He\,{\sc i} lines mainly unaffected. 

The ratios of the EWs of the high-excitation emission
lines and the EWs of the nearby hydrogen Balmer lines could provide
information on relative quantities of the correspondingly ionized atoms. 
Fig. \ref{fig6} shows that these ratios from quiescence and hot
outbursts occupy essentially the same region of the plot, only the scatter of
the outburst points is somewhat larger. Data points from the 
cool outburst (squares) are located separately. They mostly occupy the lower left
section of the plot. This result confirms that the number of the highly ionized
He and O atoms decreased significantly during the cool outburst.

\subsection{Orbital variations}

\begin{figure}
\includegraphics[width=84mm, angle=0]{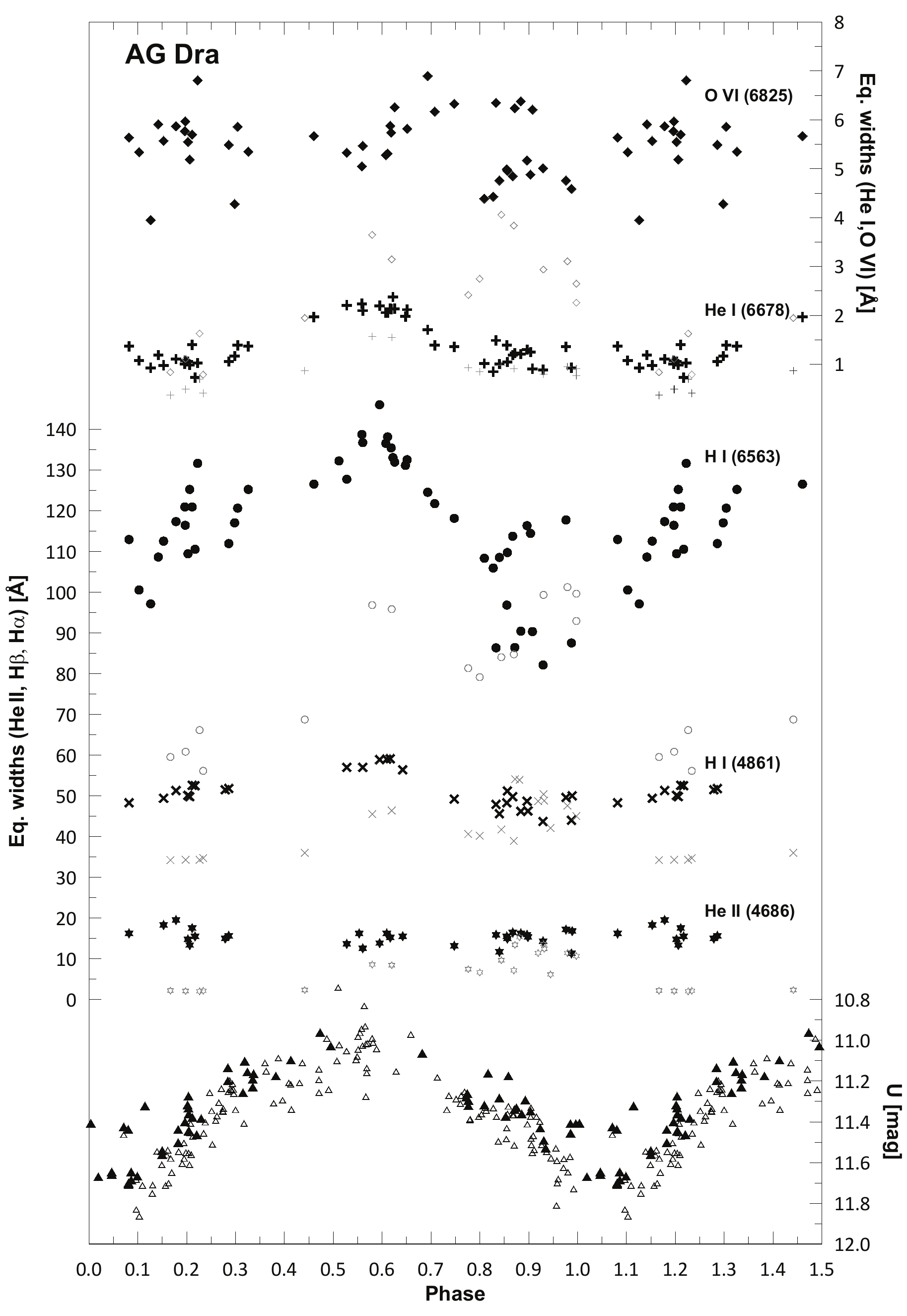}
\null
\vspace{3mm}
\caption{Dependence of the emission line EWs and the $U$ brightness on
the orbital phase. Full and black symbols correspond to the possible
quiescent episode between active stages E and F (JD\,2\,451\,200 --
2\,452\,100). Empty and/or grey symbols correspond to the quiescent stage
Q6 (after JD\,2\,454\,550).}
\label{fig7}
\end{figure}

It has long been debated whether the spectral lines of \mbox{AG Dra} vary
regularly with orbital phase. From what is said
above about the outburst variability of the emission lines, it is clear that
such regularity should only be searched for in the quiescence data. Fig.
\ref{fig7} shows the emission line EWs in quiescence 
phased to the orbital period $P = 549\fd73$ \citep{gal99} together with 
the $U$ light curve. In the latter,
the well known wave-like variability mentioned already by \citet{mein79} is
clearly seen. A very similar variability can be detected in the EWs of 
H\,$\alpha$ line, and to a lesser extent in H\,$\beta$ and He\,{\sc i}
$\lambda$ 6678 line. These low excitation lines most likely arise in an
extended gaseous volume which also emits continuum radiation in the near-UV
and optical spectral region. The high excitation He\,{\sc
ii} $\lambda$ 4686 line practically does not vary with orbital motion. This
line should have its origin close to the hot component. The EWs of the other high
excitation line, O\,{\sc vi} $\lambda$ 6825, exhibit a large scatter because
both the vicinity of the hot component and the
extended volume of the neutral hydrogen are involved in the formation of this line.
A possible relation of the occurrence of double-peaked H\,$\alpha$
profiles with the orbital motion requires further study and, thus, remains beyond the scope of this paper.

\section{Comparison with other symbiotic stars}
\label{sect4}

Symbiotic stars display a great variety of outbursts and active
stages. A few symbiotic novae \citep{ken86, belcz00} 
(like \mbox{AG Peg}, \mbox{V1016 Cyg}, \mbox{PU Vul}, \mbox{RR Tel} etc.)
have shown
only one large amplitude outburst during the observational
history. A few symbiotic recurrent novae
\citep{ken86, anu13} undergo outbursts with amplitudes up to
$ \sim 7$ magnitudes with a recurrence time of a few decades (\mbox{RS Oph} and \mbox{T
CrB} are the best known examples).
Some symbiotics are quiescent with no recorded outbursts (\mbox{RW
Hya}, \mbox{EG And} etc.), 
but the most
widespread type of activity among symbiotic stars are the so-called
classical or \mbox{Z And} type outbursts, which have amplitudes of
about $1-3$ magnitudes at optical wavelengths and recur in intervals of  
about one or a few years (e.g. \mbox{AG Dra}, \mbox{Z And}, \mbox{BF Cyg}, \mbox{CI
Cyg}) or $10-15$ years (e.g. \mbox{AX Per}) (AAVSO database, http://www.aavso.org).

\begin{figure}
\includegraphics[width=84mm,angle=0]{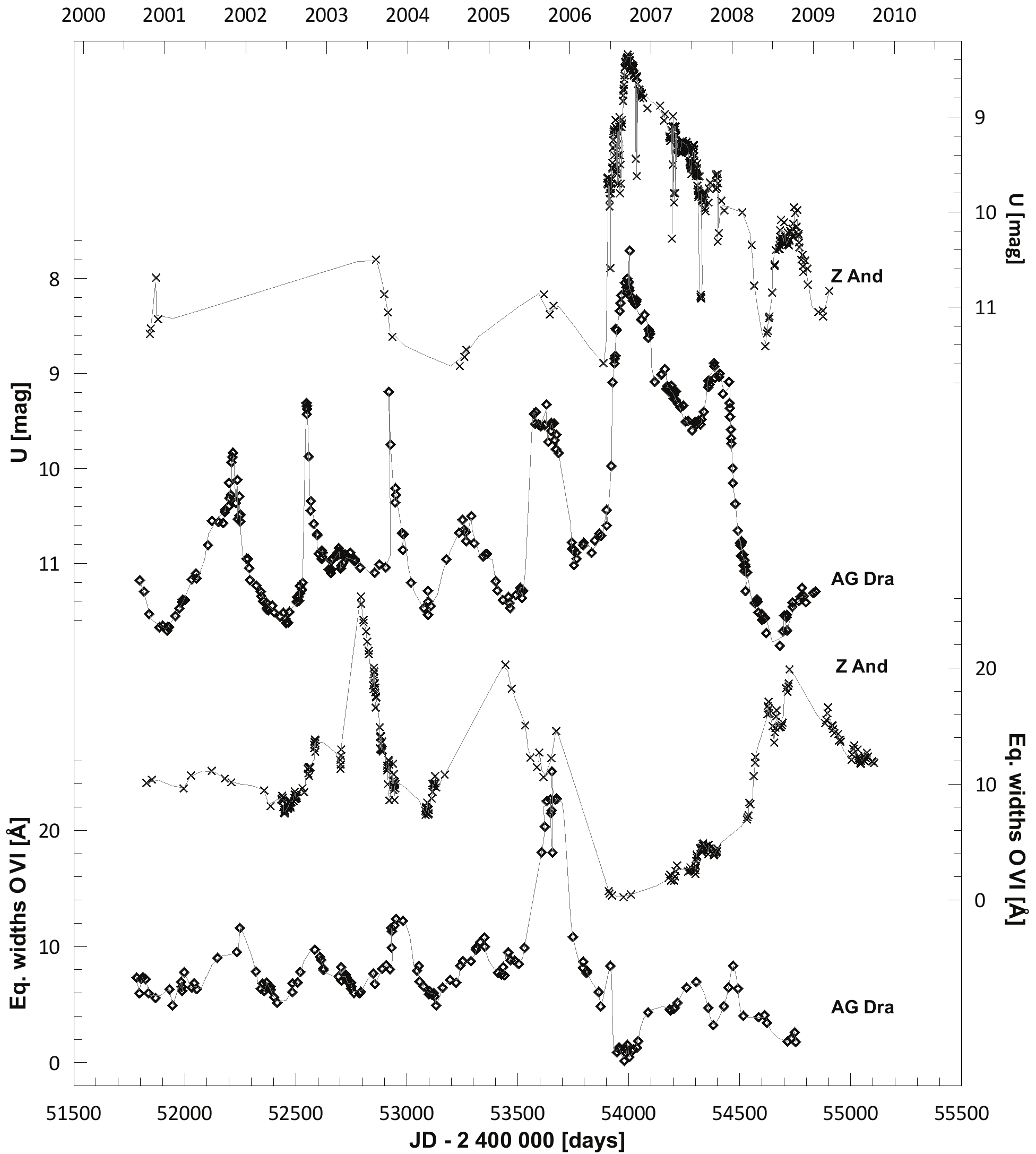}
\null
\vspace{3mm}
\caption{The $U$ light curves (top) and EWs of the \mbox{O\,{\sc vi}} 
line
at $\lambda$ 6825 (bottom) of \mbox{AG Dra} and \mbox{Z And}. The 
curves
of \mbox{Z And} are time-shifted by 2108.7 and 2102.7 days for photometric
and spectroscopic data, respectively.}
\label{fig8}
\end{figure}

A characteristic feature of some such outbursts is the weakening or
disappearance of the high excitation emission lines, mentioned 
by \citet{ken86}, and followed in more detail in \mbox{Hen 3-1341}
\citep{mun05}, \mbox{Z And} \citep{soko06, burleed07}, and \mbox{AG Dra}
\citep[][and present paper]{mun09, sho10}. As recent outbursts of the
prototypical symbiotic star \mbox{Z And} have been extensively studied, we were
tempted to quantitatively compare the behaviour of
\mbox{AG Dra} and \mbox{Z And}.
Extensive photometric time series for \mbox{Z And} are available in
\citet{skop98, skop00, skop02, skop04, skop07, soko06, tom04}.
Using these data, we selected time intervals of about 3\,000 days from the
$U$, $B$ and $V$ light curves of both \mbox{AG Dra} and \mbox{Z And},
and performed a cross-correlation analysis. The results of such
analysis for curves with complex behaviour depend on
the selection of a particular time interval.
Therefore, we performed around 20 runs with different starting times and
lengths of the analysed time intervals (1\,520 - 3\,066 days). In all
runs, the cross-correlation functions manifested the significant maxima
only for the time shift with a value around 2\,110 days. The time shifts for
the maxima of the cross-correlation functions, the corresponding maximum values
of these functions, and their errors for the light curves in the $U$, $B$ and 
$V$
filters are listed in Table \ref{table2}. Such significant correlations might not be so
surprising as we know that the $U$, $B$ and $V$ light curves of \mbox{AG Dra}
correlate well between themselves during active stages \citep{hric14}, and
the same can be concluded from the visual inspection of the light curves of
\mbox{Z And}.

It would be more intriguing to find similar correlations in some
spectroscopic features. \citet{soko06} have published the EWs
of the most prominent emission lines of Z And. We
selected the \mbox{O\,{\sc vi}} line at $\lambda$ 6825 and performed a
similar cross-correlation analysis of its EWs in \mbox{AG Dra} and
\mbox{Z And}. In all runs, the maximum value of the cross-correlation
function appeared for the time shift around 2\,290 days. Moreover, in
many runs this maximum has a double-peaked structure with the second
maximum around 2\,100 days. The maximum for time shift with value around
2\,970 days is also present in less than a half of the runs. The 
particular
results of this analysis are listed in Table \ref{table2}. The large value
of the mean error for the time shift 2\,289.6 days is probably caused by the double-peaked structure of the
maxima of the cross-correlation function.

\begin{table}
\caption{The results of the cross-correlation analysis of the $U$, $B$
and $V$ light curves as well as the EW curves of the \mbox{O\,{\sc vi}} line
at $\lambda$ 6825 of \mbox{AG Dra} and \mbox{Z And}. The $2\,\sigma$ white
level noise was about 0.1 and 0.16 for the photometric and spectroscopic data,
respectively.}
\label{table2}
\centering
\begin{tabular}{r c c c c}
\hline
Data set & \multicolumn{2}{c}{Time shift} & 
\multicolumn{2}{c}{Cross-corr. function}\\
          & median & error                & median & 
error                          \\
\hline
Filter $U$ & 2\,108.7 & 8.2  & 0.76 & 0.05 \\
Filter $B$ & 2\,108.7 & 5.7  & 0.84 & 0.03 \\
Filter $V$ & 2\,116.3 & 3.2  & 0.82 & 0.04 \\
EWs        & 2\,289.8 & 24.7 & 0.59 & 0.05 \\
            & 2\,102.7 & 4.2  & 0.49 & 0.08 \\
            & 2\,972.3 & 6.4  & 0.31 & 0.03 \\
\hline
\end{tabular}
\end{table}

These results show strong similarity of the photometric and spectroscopic
behaviour of \mbox{AG Dra} and \mbox{Z And}. The difference between the
time shift for the light and the EW curves (around 180 days) is not
essential,
because the value of the time shift for the EW curves lies indeed in the 
range
2\,100 - 2\,300 days, and its particular value depends only on the selection
of the time interval. The correspondingly 
shifted
$U$ light curves and EW variability curves of \mbox{AG Dra} and \mbox{Z 
And}
are shown in Fig. \ref{fig8}.

Those reasonable correlations could imply some similarity of the nature
of the hot components and the mechanisms of the outbursts in \mbox{AG Dra} and
\mbox{Z And}. Cool giants of these symbiotic stars are of rather different
spectral types, K2--K3 and M4--M5,
respectively. Available estimates of the mass of the hot component (WD) are
similar for both stars, $0.4-0.6 M_{\sun}$ (with a preference towards a higher
value) for AG Dra \citep{mik95}, and $0.65 \pm 0.28 M_{\sun}$ for
\mbox{Z And} \citep{sch97}. \citet{fekel00} have adopted $0.5 M_{\sun}$ for
\mbox{AG Dra}.
Accordingly, the physical characteristics of the outbursts could indeed be
similar in both stars, provided that the mass accretion rates onto the WD are
comparable. The lower mass loss rate of the K giant compared to the M giant
could be compensated by the smaller separation of the components in \mbox{AG
Dra}.

\section{Discussion and conclusions}
\label{sect5}

It has been established long ago that a white dwarf alone is not
capable of ionizing the nebulae of symbiotic stars to
produce the characteristic emission-line spectrum. 
Quasi-steady thermonuclear shell burning on the surface of the
WD can provide additional energy, and this phenomenon appears
to be common in most of the symbiotic systems \citep{paczy78,
iben82, siore92}. An accretion rate of a few times $10^{-8}
M_{\sun} \rmn{yr}^{-1}$ is sufficient to produce a luminosity
of $10^3 L_{\sun}$ by a thermonuclear shell burning
\citep{soko06}.
The classical symbiotic outbursts, recurring every few years or
decades, however, do not fit well into this picture.
They are by far too frequent to be
nova-like thermonuclear runaways like those in symbiotic
recurrent novae \citep{iben82, kentru83, mik95}. Even with the
WD mass close to the Chandrasekhar limit, the recurrence times
would be tens or hundreds of years \citep{sion79}. But WDs in
symbiotic systems seem to have a low mass (around $0.6 M_{\sun}$) in
most cases \citep{mik95}.

The "combination nova" model proposed by \citet{soko06} seems to be a
promising explanation, at least to the behaviour of {\mbox Z And}. Smaller scale hot
outbursts (e.g. in 1997) are explained by the accretion disc
instability model like in dwarf novae \citep{warn95}. The $2000-2002$
outburst of \mbox{Z And} was similar to the major (cool) outbursts of \mbox{AG
Dra}.
According to \citet{soko06}, the disc instability in this case triggered
enhanced thermonuclear shell burning on the WD surface, thus resembling
a classical nova outburst. For the $0.65\,M_{\sun}$ WD, an average accretion
rate of $5 \cdot 10^{-8} M_{\sun} \rmn{yr}^{-1}$ would be enough to both fuel
the quasi-steady nuclear burning and accumulate enough material in the disc
to trigger a combination nova event on time-scales of 10 years. The major (cool)
outbursts of \mbox{AG Dra} occur on a similar time-scale ($12-15$ years). However,
there is a fundamental difficulty because we do not have a firm evidence for the
presence of an accretion disc in \mbox{AG Dra}. Radio observations of \mbox{AG
Dra}
have indicated bipolar outflows \citep{ogl02, mik02}, but those can be
explained hydrodynamically by a disc-like density condensation in the
orbital plane \citep{gawr03}. No evidence of collimated bipolar jets
like in \mbox{Z And} \citep{burleed07} neither short
time-scale flickering \citep{soko01} has been found in \mbox{AG Dra}. 

A "combination nova" model might be applicable to \mbox{AG
Dra}, provided that a trigger mechanism for thermonuclear shell flashes can
be found. 
Triggering of major (cool) outbursts seems to take place
close to some critical threshold, as in some cases (active stages A and C
in \citet{hric14})
there has been not enough power to ignite additional nuclear reactions
required to initiate a cool outburst.

Our conclusions can be shortly summarized as follows:

-- cool and hot outbursts of \mbox{AG Dra} can be clearly distinguished by the 
behaviour of the emission lines in the optical spectrum;

-- the Raman scattered O\,{\sc vi} line $\lambda$ 6825 almost disappeared during
the cool outburst confirming a drop in the hot component's temperature as
was also found
from the variations of other emission lines;

-- emission lines of hydrogen and neutral helium did not change
significantly during the cool outburst. They are correlated with the orbital
motion in quiescence and become stronger in hot outbursts;

-- the similarity of the outbursts of \mbox{AG Dra} and \mbox{Z And} shows that a "combination
nova" model might explain the outbursts of \mbox{AG Dra}. The presence of an accretion
disc in the system still lacks confirmation.

\section*{Acknowledgements}
This study was supported by the
Estonian Ministry of Education and Research target financed
research topic SF0060030s08 and institutional research funding IUT 40-1, by the Slovak Academy of Sciences
VEGA Grant No. 2/0038/13 and by the realisation of the Project
ITMS No. 26220120009. The authors are thankful to colleagues Kalju Annuk and
Alar Puss from Tartu Observatory who have recorded many spectra of \mbox{AG
Dra} during their observing nights on the 1.5-metre telescope, and Michaela
Kraus for a careful reading of the manuscript.  

\bsp
\label{lastpage}

\end{document}